\documentclass[12pt,preprint]{aastex}

\newcommand{\hii}{\hbox{H\,\textsc{ii}}}
\newcommand{\vsad}{\textsc{vsad}}
\newcommand{\imfit}{\textsc{imfit}}
\newcommand{\aips}{\textsc{aips}}
\newcommand{\miriad}{\textsc{miriad}}
\newcommand{\pegase}{\textsc{pegase-ii}}
\newcommand{\radiolum}{$\rm{L}_{1.4}$}
\newcommand{\uvlum}{$\rm{L}_{uv}$}
\newcommand{\halum}{$\rm{L}_{H\alpha}$}

\shorttitle{A Comparison of Star Formation Diagnostics}

\shortauthors{Sullivan et al.}

\begin{document}

\title{A Comparison of Independent Star Formation Diagnostics for a
UV-Selected Sample of Nearby Galaxies}

\author{Mark~Sullivan\altaffilmark{1,2},
  Bahram~Mobasher\altaffilmark{3,4}, Ben~Chan\altaffilmark{5},
  Lawrence~Cram\altaffilmark{5}, Richard~Ellis\altaffilmark{2},
  Marie~Treyer\altaffilmark{6}, Andrew~Hopkins\altaffilmark{7}}

\altaffiltext{1}{Institute of Astronomy, Madingley Road, Cambridge CB3 OHA, UK}
\altaffiltext{2}{California Institute of Technology, E. California Blvd, Pasadena CA 91125, USA}
\altaffiltext{3}{Space Telescope Science Institute, 3700 San Martin Drive, Baltimore, MD 21218, USA}
\altaffiltext{4}{Affiliated with the Astrophysics Division of the European Space Agency}
\altaffiltext{5}{School of Physics, University of Sydney, Sydney NSW 2006, Australia}
\altaffiltext{6}{Laboratoire d'Astronomie Spatiale, Traverse du Siphon, 19976 Marseille, France}
\altaffiltext{7}{Department of Physics and Astronomy, University of Pittsburgh, 3941 O'Hara Street, Pittsburgh, PA 15260, USA}

\begin{abstract}
  We present results from a decimetric radio survey undertaken with
  the Very Large Array (VLA) as part of a longer term goal to
  inter-compare star formation and dust extinction diagnostics, on a
  galaxy by galaxy basis, for a representative sample of nearby
  galaxies. For our survey field, Selected Area 57, star formation
  rates derived from 1.4\,GHz luminosities are compared with earlier
  nebular emission line and ultraviolet (UV) continuum diagnostics.
  We find broad correlations, over several decades in luminosity,
  between H$\alpha$, the UV continuum and 1.4\,GHz diagnostics.
  However, the scatter in these relations is found to be larger than
  observational errors, with offsets between the observed relations
  and those expected assuming constant star-formation histories and
  luminosity-independent extinction models. We investigate the
  physical origin of the observed relations, and conclude the
  discrepancies between different star-formation diagnostics can only
  be partly explained by simple models of dust extinction in galaxies.
  These models cannot by themselves explain all the observed
  differences, introducing the need for temporally varying
  star-formation histories and/or more complex models of extinction,
  to explain the entire dataset.

\end{abstract}


\section{Introduction}

Considerable observational progress has been made in tracking the star
formation rate (SFR) per comoving volume element as a function of
redshift \citep[see][]{1996MNRAS.283.1388M}. The form of this ``cosmic
star formation history'' is important not only in indicating likely
eras of dominant activity but also in securing a self-consistent
picture of chemical enrichment for detailed comparison with
intergalactic absorption line diagnostics
\citep*{1999ApJ...522..604P}, as well as with the predictions of
popular semi-analytical models of galaxy formation
\citep[e.g.][]{1998ApJ...498..504B,2000MNRAS.319..168C}.  Various
observational diagnostics have been employed to determine the SFR of a
galaxy in a particular redshift survey. These include (i) Hydrogen
series (i.e. Balmer or Lyman) or other nebular emission lines,
generated in regions ionised by the most massive ($\gtrsim
10\,\rm{M}_{\odot}$) early-type stars
\citep{1995ApJ...455L...1G,1998ApJ...495..691T,1999MNRAS.306..843G,2000A&A...362....9M},
(ii) the ultraviolet (UV) continuum, dominated by young (but less
massive, $\gtrsim 5\,\rm{M}_{\odot}$) stars
\citep{1996ApJ...460L...1L,1996MNRAS.283.1388M,1998MNRAS.300..303T,1999AJ....118..603C,1999ApJ...519....1S,2000MNRAS.312..442S},
(iii) far infrared (FIR) luminosities arising from thermal emission
from dust absorbed UV radiation from the more massive stars
\citep{1997MNRAS.289..490R,1999MNRAS.302..632B}, and (iv) 1.4\,GHz
radio emission, thought to originate mainly via synchrotron radiation
generated by relativistic electrons accelerated by type II supernovae
(SNe) from stars of mass $\gtrsim 7\,-\,8\,\rm{M}_{\odot}$
\citep{1992ARA&A..30..575C,1998ApJ...507..155C}

Each diagnostic has its own merits, disadvantages and uncertainties
\citep[for a review see][]{1998ARA&A..36..189K} and none can yet be
applied uniformly to samples over the full observed redshift range.
Accordingly, it is inevitable that the cosmic star formation history
(SFH) has been pieced together using different diagnostics from
independent samples of galaxies. Therefore, it is crucial to know how
well the various diagnostics agree {\em on a galaxy by galaxy basis}.
A major source of uncertainty is the likelihood that some diagnostics
are affected by the presence of obscuring dust. Radio continuum
diagnostics have the distinct advantage of being immune from dust
extinction. Non-uniformities in the SFH (or a time varying SFR) are
also important, and might, if present, introduce additional biases
between the different SFR diagnostics.

Only limited work has been done inter-comparing different
star-formation (SF) diagnostics for galaxies, with studies at low and
intermediate redshifts showing significant discrepancies.
\citet{1998ApJ...507..155C} compared the SFRs from H$\alpha$, FIR and
$U$-band measurements with those from decimetric radio luminosity
(\radiolum) for a sample of over 700 local galaxies. Though the
various SFR estimates were in broad agreement, numerous systematic
differences were found, including a significant scatter in all the
relationships compared to the \radiolum/FIR relation. They conclude
that the scatter may partly be due to sample selection (i.e.
inhomogeneities in the sample selection or possible AGN contamination)
or have a real physical basis (e.g. differential extinction among
individual galaxies or time-dependent initial mass functions). Studies
comparing UV continuum and H$\alpha$ emission fluxes likewise reveal
systematic discrepancies and a large scatter
\citep{1999MNRAS.306..843G,2000MNRAS.312..442S,2001ApJ...548..681B}.
Although extinction is a likely source of the scatter,
\citet[hereafter S2000]{2000MNRAS.312..442S} argued that the observed
trend on the UV-H$\alpha$ plane is consistent with non-uniform SFHs for
some fraction of the population. If starburst activities are a common
occurrence for high redshift galaxies as some models suggest
\citep*{2001MNRAS.320..504S}, further corrections may be needed to
derive representative SFRs from optical and UV diagnostics based on
luminous sources only.

This paper is motivated by the need for more comprehensive and
rigorous investigation into the inter-relationships between the
various SF diagnostics, including those based on the promising
1.4\,GHz luminosity. Though the radio emission is generally thought to
be generated as a by-product of the supernovae of massive stars (and
therefore a measure of SFR), the steps relating the supernovae
explosion to the arrival of radio emission at the Earth -- for example
the electron acceleration in the supernova remnant and the subsequent
cosmic ray propagation, the role of magnetic fields, any possible
energy loss mechanisms, or a significant cosmic-ray escape fraction --
are not yet well understood. We are therefore particularly interested
in the quantitative empirical precision with which radio luminosities
can be used to trace SF. The major hurdle is the need for a large,
well-controlled sample with considerable overlap in the various
diagnostics.  For this study, we choose the UV-selected sample
discussed by \citet{1998MNRAS.300..303T} and S2000, drawn from the
FOCA balloon-borne imaging data of Selected Area 57 (SA57), taken by
\citet{1992A&A...257...24M}. The SA57 survey contains 222
spectroscopically-confirmed star-forming galaxies, forming a
statistically-complete sample selected at 2000\,{\AA}.

A plan of the paper follows. In ${\S}$2 we review the SA57 dataset and
briefly describe the 1.4\,GHz observations (a more detailed
description will be given by Chan et al., in preparation). In ${\S}$3
we discuss the adopted flux-SFR calibrations, the selection effects
present in our sample, and inter-compare the SFRs derived from the
different diagnostics. In ${\S}$4 we discuss the implications and
present our conclusions. Throughout we assume $\Omega_{\rm{M}}=1$,
$H_0=100\,\rm{km\,s^{-1}\,Mpc^{-1}}$, solar metallicity, and, unless
otherwise stated, our adopted initial mass function is
\citet{1955ApJ...121..161S}-like ($\Psi(M) \sim M^{-2.35}$) with mass
limits $\rm{M}_{\rm{lower}}=0.1$ and
$\rm{M}_{\rm{upper}}=100\,\rm{M_{\odot}}$.

\section{The Datasets}
\label{data_section}

The UV-selected sample is drawn from the \citet{1992A&A...257...24M}
FOCA\,1500 1.55$^{\circ}$ diameter image of Selected Area 57 centered
at $\rm{RA}=13^{\rm{h}}03^{\rm{m}}53^{\rm{s}}$, $\rm{Dec.}=+29\degr
20\arcmin 30\arcsec$ (B1950). The photometric details are presented in
\citet{1998MNRAS.300..303T} and S2000.  Briefly, the errors in the UV
photometry range from 0.2\,mag at the bright end to 0.5\,mag at the
fainter end. Optical fiber spectroscopy has been secured for an
unbiased (i.e. randomly selected) subset of 369 UV sources limited at
$m_{UV}=18.5$ (equivalent to $B\simeq 20-21.5$ for a late-type galaxy)
using the WIYN observatory\footnote[1]{The WIYN Observatory is a joint
  facility of the University of Wisconsin-Madison, Indiana University,
  Yale University, and the National Optical Astronomy Observatories.}
($\lambda\lambda=3500-6600\,\mbox{\AA}$) and the William Herschel
Telescope (WHT, $\lambda\lambda=3500-9000\,\mbox{\AA}$). This
spectroscopy provides redshifts for 222 UV-selected galaxies, of which
178 were found to have strong emission line spectra.  However, only
the WHT spectra have sufficient wavelength coverage to detect
H$\alpha$, and only for a subsample of 88 WHT-observed galaxies is the
flux calibration sufficient for a reliable conversion of H$\alpha$ to
SFR (note this does \textit{not} include all the galaxies with
detected H$\alpha$ emission).  The errors in the H$\alpha$ fluxes are
estimated from the S/N of the spectrum in question. A fiber size of
2.7'' diameter was used for the flux-calibrated optical spectroscopy;
no aperture corrections are applied, hence the H$\alpha$ flux is
underestimated in very nearby objects with a larger apparent size (see
S2000).

The 1.4\,GHz radio observations were carried out at the NRAO VLA
telescope, and cover the central square degree of the SA57 field. A
$4\,\sigma$ detection limit of $170\,\mu\rm{Jy}$ was achieved at the
centre of the field dropping to $340\,\mu\rm{Jy}$ in the outer
regions. A detailed discussion of the VLA radio data and its reduction
will be presented in Chan et al. (in prep.). Briefly, the task is to
measure radio flux densities, or to estimate upper flux limits, at the
positions of all the FOCA-UV galaxies. This problem is not the same as
deriving a catalogue from the image, since the search positions, at
which the FOCA galaxies are located, are pre-determined. The
position-dependent noise in the image, arising from the primary beam
response and the mosaicing method, complicates this problem. We begin
by constructing an image of the position-dependent noise, using the
measured noise near the centers of the pointings in conjunction with
the primary beam attenuation and the overlapping of the mosaic. We
then search near each UV position. For radio sources with flux
densities greater than $4\sigma$, we use the \aips\ task \vsad\ to
determine the source properties from a one-component Gaussian fit in
which the amplitude, size and position of the source is allowed to
vary. Three of these fits reveal very extended sources, whose
properties were recalculated from the individual pixel values.  For
sources fainter than $4\sigma$, a Gaussian fit was made with the
position fixed at the UV source position using the \miriad\ task
\imfit\ (as \vsad\ does not allow position-fixed fitting). If the
amplitude of this fit exceeds the local $1\sigma$ noise level, it is
reported as a detection, while fainter fits are reported as upper
limits using the $1\sigma$ level.  For all reported fits, the
$1\sigma$ error is reported based on the noise image.

A total of 26 out of 191 FOCA galaxies (with a secure redshift) within
the central 1.4\,GHz survey area of SA57 are detected using \vsad\ at
a significance level of $\geq 4\,\sigma$.  Of these, 22 have single
optical counterparts on the Palomar Sky Survey (POSS) plates (see
S2000 for a discussion of FOCA/POSS identification procedures). A
further three FOCA sources were identified with extended radio
emission that provided poor fits to a Gaussian profile; the fluxes for
these were consequently calculated by adding together all the pixels
inside the extended emission area as described above.  The position
fixed fitting identifies a further 25 galaxies, 18 with unambiguous
optical counterparts.  Since in the present paper we only wish to
study field galaxies (avoiding contamination by known cluster members
which are likely to reside in different environments), we also exclude
3 galaxies in the redshift range of the nearby Coma cluster
($0.020<z<0.027$). A summary of the sample numbers is given in
Table~\ref{sample_numbers}. Of the final sample, 25 WHT-observed
galaxy spectra have detected H$\alpha$ of which 17 have an adequate
flux calibration to derive SFRs.

All observed radio luminosities have been $k$-corrected to
luminosities appropriate to a rest-frame frequency of 1.4\,GHz
according to $L_{1.4}^{0}=L_{1.4}^{obs}\times(1+z)^{-0.8}$, where the
spectral index of -0.8 is typical for the non-thermal synchrotron
component of decimetric radiation.

\section{Inter-comparison of star formation diagnostics}
\label{sec:interc-star-form}

\subsection{Diagnostic Relations}
\label{sec:diagnostic-relations}

To compare the SF diagnostics based on the radio (\radiolum), UV
(\uvlum), and H$\alpha$ (\halum) luminosities, a self-consistent
calibration is required. This is accomplished using the \pegase\ 
spectral synthesis code \citep{1999astro.ph.9912179} which generates
galaxy spectra as a function of time for arbitrary SFHs, from which
the H$\alpha$ and UV luminosities can be calculated (S2000). In order
to estimate 1.4\,GHz luminosities for a given SF scenario, we add a
simple prescription using the type II supernovae rate given by
\pegase, using calibrations from \citet{1992ARA&A..30..575C}. We
calculate the conversion factors for four of the different
metallicities available in \pegase; we report solar metallicity
conversions below, and the remainder in Table~\ref{conversion_factors}.
Assuming a constant SFH, the three SF diagnostics in this study are
calibrated as explained below:

\noindent \textit{H$\alpha$ emission:} The relevant photons originate
via re-processed ionizing radiation at wavelengths $\lambda<912$\,{\AA}\ 
produced by the most massive ($>10\,\rm{M}_{\odot}$), short-lived
($\simeq20\,\rm{Myr}$), OB stars.  Accordingly, H$\alpha$ emission is a
virtually instantaneous SF measure -- for a burst of star-formation
with a constant SFR, the H$\alpha$ emission can be considered constant
after $\simeq10\,\rm{Myr}$. However, as it depends so strongly on the
most massive stars, it is very sensitive to the form of the initial
mass function (IMF) \citep[see][for example]{1998ARA&A..36..189K}.
There is much debate in the literature as to the universality, or
otherwise, of the IMF \citep[see][for recent
reviews]{1998simf.conf..201S,2001astro.ph..0102189,2001astro.ph..0101321};
here we assume the form to be universal, but discuss variations in the
upper mass cut-off in Section~\ref{sec:phys-orig-scatt}.  Most
calibrations assume case-B recombination \citep[a comprehensive
treatment of its use can be found in][]{2001astro.ph..1097C}. Using
\pegase\ we derive:

\begin{equation}
SFR=\frac{L_{H\alpha} (\rm{erg\,s^{-1}})}{1.22\times10^{41}}\;\rm{M_{\odot}}\;\rm{yr^{-1}}.
\end{equation}

\noindent
As a comparison, \citet{1998ARA&A..36..189K} derive a conversion value
of $1.26\times10^{41}$, very close to the value used here.

\noindent
\textit{UV 2000\,{\AA}\ continuum:} As stars that contribute to
radiation at 2000\,{\AA}\ span a range of ages (and hence initial
masses), including some post-main sequence contribution, any UV-SFR
calibration is dependent on the past history of SF.  This introduces a
significant uncertainty when interpreting the observations of the
star-forming galaxies in this sample. As a guide, we calculate
conversion values based on constant burst of SF of durations  10,
100 and 1000\,Myr using \pegase. This gives:

\begin{equation}
SFR=\frac{L_{UV} (\rm{erg\,s^{-1}\,\mbox{\AA}^{-1}})}{x\times10^{39}}\;\rm{M_{\odot}}\;\rm{yr^{-1}}.
\end{equation}

\noindent 
where $x$ is equal to 3.75, 5.76 and 6.56 for 10, 100 and 1000\,Myr
respectively. \citet{1998ARA&A..36..189K} derive a conversion value of
$5.36\times10^{39}$, again in good agreement.

\noindent
\textit{1.4\,GHz continuum:} The integrated radio flux at 1.4\,GHz is
thought to originate via non-thermal synchrotron emission from
electrons accelerated by supernovae (spectral index $\simeq-0.8$) and
thermal Bremsstrahlung from electrons in ionised \hii\ regions
(spectral index $\simeq-0.1$). Non-thermal emission dominates (around
90\%) at the frequency of interest. We introduce a prescription into
\pegase\ to predict the non-thermal 1.4\,GHz luminosity, \radiolum,
based on the radio supernova rate (type II SNe) in the \pegase\ code.
For a constant SFR burst, the SNe\,II rate is effectively constant
after $\simeq80\,\rm{Myr}$. Using the relationship between non-thermal
radio luminosity and the radio supernova rate, presented in
\citet{1990ApJ...357...97C} \citep[see also][]{1992ARA&A..30..575C},
the SFR as a function of 1.4\,GHz luminosity for our assumed IMF is:

\begin{equation}
SFR=\frac{L_{1.4}(\rm{erg\,s^{-1}\,Hz^{-1}})}{x\times10^{27}} \;\rm{M_{\odot}}\;\rm{yr^{-1}}.
\end{equation}

\noindent
where $x$ is equal to 2.39 and 8.85 for 10 and 100\,Myr respectively.
This agrees well with a conversion factor of $7.36\times10^{27}$ given
by \citet{2000ApJ...544..641H}, which maps non-thermal luminosity to
SFR at 1.4\,GHz. The small difference arises as \pegase\ calculates a
smaller (metallicity dependent) lower mass cut-off for type II SNe
than the $8\,\rm{M}_{\odot}$ used in \citet{1992ARA&A..30..575C}, as
convective overshooting increases the mass of the degenerate core, and
the Chandrasekhar mass is reached more easily (M.~Fioc~2000, private
communication), hence the conversion factor above is slightly
metallicity dependent (see Table~\ref{conversion_factors}). We neglect
the small amount of thermal emission at 1.4\,GHz.

There is some evidence to suggest that the calibration of
\radiolum\,--\,SFR may not be perfectly linear, particularly in low
luminosity (or low SFR) objects where there is the possibility that a
fraction of the SN remnant-accelerated cosmic rays may escape from the
galaxy \citep*{1991ApJ...376...95C}. In such scenarios, the true SFR
for a galaxy may be underestimated when using the linear relationship
presented above.

Our `standard' scenario then uses the calibrations listed above, and
the following implicit assumptions: i) constant SFHs over recent
timescales for the galaxies under study, ii) a Salpeter IMF, iii)
solar metallicity and the stellar libraries/evolutionary tracks used
in \pegase\ \citep{1999astro.ph.9912179}, and iv) a simple, tight and linear
relationship between \radiolum\ and SFR. We will discuss the validity
of some of these assumptions in later sections.

\subsection{The effects of selection bias}
\label{sec:effects-select-bias}

Before analysing the present dataset in more detail, it is important
to explore possible selection effects that might operate in any
comparison where only a subset of one diagnostic set is used. In
particular, we need to determine that any correlations seen in our
data are not merely products of the potentially complex selection
effects in this study.  In this section we attempt to identify which
types of galaxies are excluded from our combined survey due to the
different flux limits, and how this might affect any correlations that
we see in our data.

Our sample has two independent flux limits. The first arises from the
original selection at 2000\,{\AA}, which was limited at $m_{uv}=18.5$,
a flux limit of
$1.35\times10^{-16}\,\rm{erg}\,\rm{s}^{-1}\,\rm{cm}^{-2}\,\mbox{\AA}^{-1}$.
The second flux limit is that of the subsequent radio follow-up
survey. We find this can be well approximated by a central
$0.20\,\rm{deg}^2$ region with a $4\,\sigma$ sensitivity of
$170\,\mu\rm{Jy}$ and an outer $0.75\,\rm{deg}^2$ region with a
sensitivity of $340\,\mu\rm{Jy}$. The subsequent spectroscopic
analysis of the UV-selected sample introduces further biases, as
firstly only objects detected in the UV have any H$\alpha$ (or
redshift) information, and secondly only objects in our sample which
lie at $z_{spec}\leq 0.4$, the redshift at which H$\alpha$ is shifted
out of our spectroscopic window, have information on this diagnostic
line.

To study these selection criteria, we predict the expected number
count distribution of our sample over the FOCA survey area as a
function of apparent UV magnitude ($m_{uv}$) and redshift (i.e.
$n(m_{uv},z)dm_{uv}dz$) as

\begin{equation}
n(m_{uv},z)dm_{uv}dz=\frac{\omega}{4\pi}\frac{dV}{dz}\phi(m_{uv},z)p(m_{uv})dm_{uv}dz
\end{equation}

\noindent
where $\phi(m_{uv},z)$ is derived from the local (dust-uncorrected) UV
luminosity function $\phi(M_{uv})$
\citep[see][S2000]{1998MNRAS.300..303T}, $p(m_{uv})$ is the
spectroscopic completeness as given in S2000 and $\omega$ is the survey
area in steradians.

For every galaxy in each $(m_{uv},z)$ bin, a variable and moderate
dust extinction at 2000\,{\AA}\ ($A_{2000}$) is then simulated,
brightening the UV magnitudes by a random amount of
$A_{2000}=0\,-\,1$\,mag. To simulate the observational uncertainties,
we apply a random ``error'' to each galaxy measurement based on the
observational errors discussed in Section~\ref{data_section}. To
convert these dust-corrected UV magnitudes to an expected 1.4\,GHz
luminosity we use the ratio of the star-forming relations derived in
Section~\ref{sec:diagnostic-relations}. As a sanity check, we also
compare this conversion value with the ratio of the characteristic
luminosities ($L_{\ast}$) for the UV and 1.4\,GHz LFs of star-forming
galaxies, taken from S2000 and \citet{1999MNRAS.308...45M}
respectively, and find agreement to within a factor of two.

\begin{figure*}
  \plotone{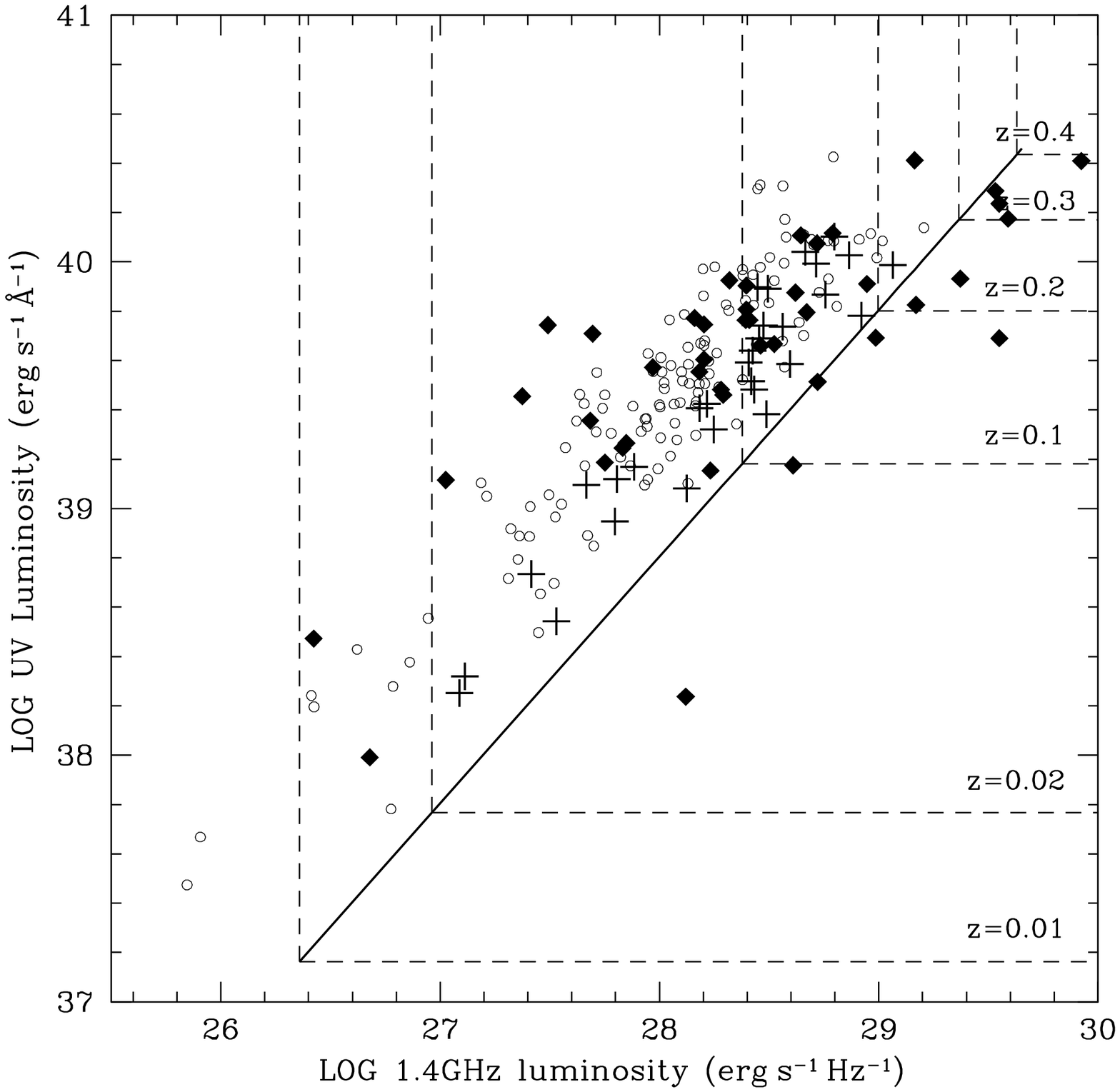} \figcaption{An
    investigation of the limitations of a joint radio and UV survey of
    star forming galaxies. Circles illustrate the likely distribution
    of a UV-selected sample according to source count and redshift
    data obtained for the $m_{UV}<18.5$ FOCA sample, a hypothetical
    \radiolum\,--\,\uvlum\ correlation, observational errors and a
    randomly distributed extinction (see text for details). Crosses
    show those UV sources that would be detected to the flux limit of
    the present VLA survey. The dashed lines indicate the formal
    regions (above and to the right), in which a galaxy at a
    particular redshift can be found based on the UV and radio survey
    limits.  Overplotted (solid diamonds) is the actual observational data.
\label{selection}}
\end{figure*}

By studying the relation between the \textit{dust-uncorrected} \uvlum\ 
and the \radiolum\ (derived from the \textit{corrected} \uvlum) the
simulation predicts the form of the relationship on the
\radiolum\,--\,\uvlum\ luminosity plane (including the effects of the
observational uncertainties) we might expect to see if our radio
survey were deep enough to detect all the FOCA galaxies
(Figure~\ref{selection}).

The simulation is then used to identify areas of the diagnostic plot
in which we do not find galaxies due to the selection effects of the
sample by excluding those objects that fall below the sensitivity of
the radio survey. The prediction from this simulation is compared in
Figure~\ref{selection} with the actual observed data.  Also shown are
the ``accessible'' areas of the diagram for galaxies at different
redshifts based on the formal flux limits of the two surveys. Clearly
we cannot detect the majority of the intrinsically low luminosity
objects in this study because of the flux limit of the radio survey,
and, for a given UV luminosity, those radio galaxies we do detect in
our simulation lie along the bottom of the distribution of model
galaxies (i.e. have brighter 1.4\,GHz luminosities).  We are thus
biased against objects at low intrinsic radio power but which are
bright enough to be detected in the UV (i.e. faint star-forming or
post-starburst objects). This can also be seen by examining the limits
by redshift, where the model galaxies lie offset from the line
denoting the respective survey limits. Extending the radio survey to
fainter flux limits will shift the vertical lines in
Figure~\ref{selection} to the left, allowing a greater fraction of the
UV galaxies to be detected at given redshift.

There is also a larger scatter in the observed data, compared to the
simulation, which assumes the \uvlum\,--\,\radiolum\ conversion to be
perfectly linear. Scatter in the simulations can be generated by
varying the adopted dust extinction on the UV luminosities over larger
ranges.  However, if we increase the range of $A_{2000}$ in this way,
we brighten the predicted radio luminosities and end up with an excess
population c.f.  the observations.  For example, 14\% of the UV
galaxies are detected at $\geq4\sigma$; with $A_{2000}=0$ the
predicted fraction of UV galaxies detected in the radio is $\sim4\%$,
for $A_{2000}=0-1$ the fraction is $\sim10\%$, for $A_{2000}=0-2$ the
fraction is $\sim20\%$ and for $A_{2000}=1-2$ the fraction is
$\sim45\%$.  Therefore, to generate a large scatter by varying the
dust extinction over larger ranges results in the detection of a
larger fraction of our UV population at 1.4\,GHz than is actually
observed. We will return to this subject in
Section~\ref{sec:phys-orig-scatt}.

In summary, the simulations in this section reflect the presence of a
bias against low power radio sources with moderate dust extinction,
which will not appear in our UV survey. However, our conclusion from
Figure~\ref{selection} is that this bias will not significantly affect
our results as it does not exclude populations of objects with
significantly different properties or luminosities to those that are
detected.

\subsection{Observational Results}
\label{sec:observ-results}

\begin{figure*}
  \plotone{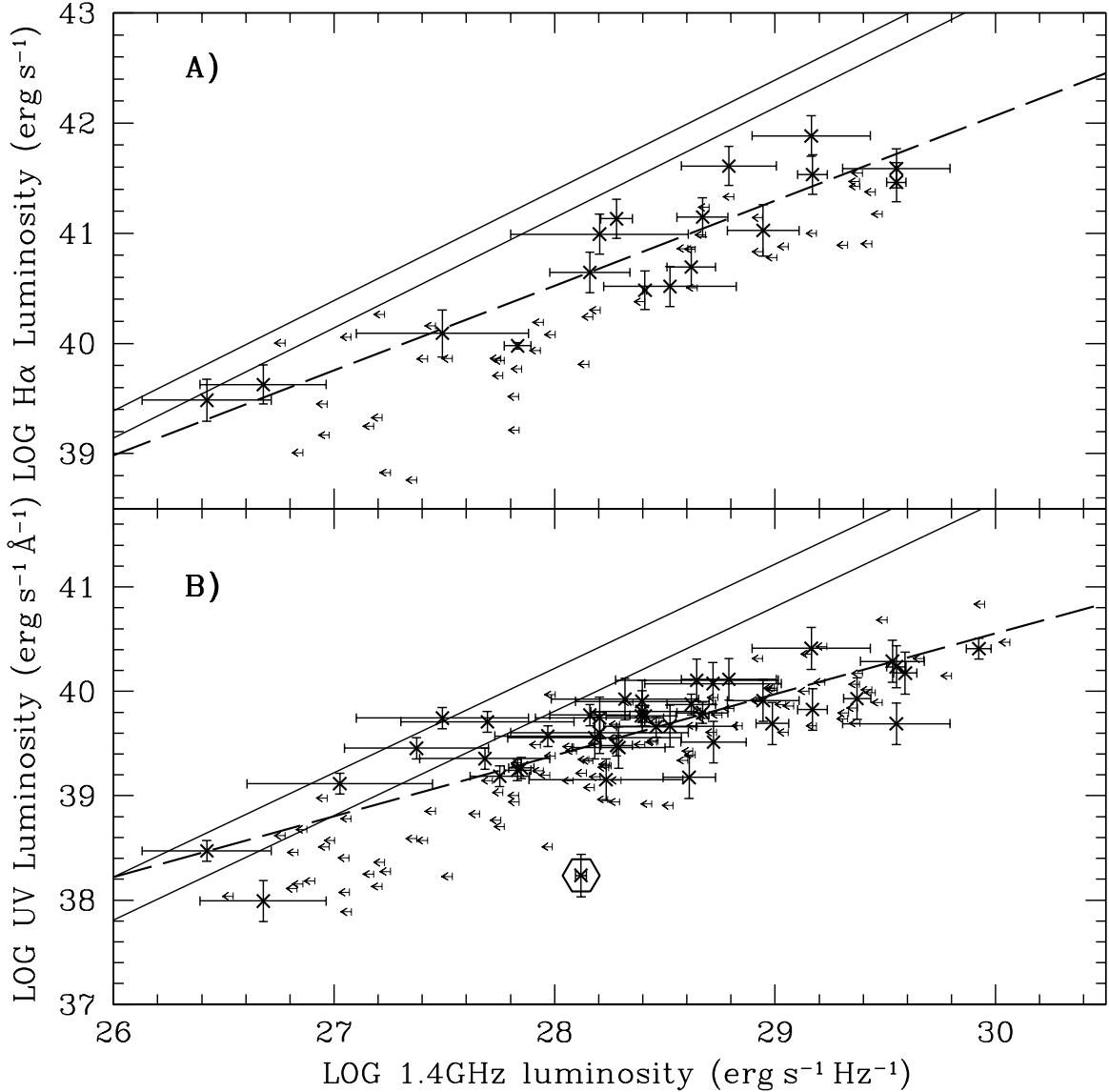} \figcaption{The correlations between the different
    SF tracers.  \textbf{A)}\,The correlation between radio and
    H$\alpha$ luminosities, \textbf{B)}\,between radio and UV
    luminosities. In both cases, the two solid lines denote equality
    of SFRs for $Z=0.02$ and $\tau=100\,\rm{Myr}$ (UPPER LINE) and
    $Z=0.004$, $tau=10\,\rm{Myr}$ (LOWER LINE), indicating the range
    of uncertainties in the luminosity-SFR conversion. Values are
    taken from Table~\ref{conversion_factors}. Both \textbf{A} and
    \textbf{B} show observed values which have not been corrected for
    dust extinction.  In both diagrams, the long-dashed lines show the
    least-squares best-fit to the galaxies. The boxed galaxy in
    \textbf{B} indicates an extremely red object compared to the other
    galaxies in the sample.
\label{correlations}}
\end{figure*}

With the above selection effects in mind, we now consider the
UV/H$\alpha$/1.4\,GHz relations. The UV, H$\alpha$, and 1.4\,GHz luminosities,
uncorrected for dust extinction, are compared in
Figure~\ref{correlations} for galaxies which have such available data.
FOCA galaxies without a reliable radio detection are shown as upper
limits in radio luminosity. Figure~\ref{correlations} demonstrates
apparent correlations between different SF diagnostics over 3\,--\,4
orders of magnitude, as predicted from
Section~\ref{sec:effects-select-bias}.

We test the strength of these correlations using various statistical
methods. We calculate the linear correlation coefficient (or
``Pearson's r''), as well as the non-parametric Spearman rank-order
correlation coefficient (see Table~\ref{fits_table}). In these tests
between any two datasets, a result of $+1$ indicates a perfect,
positive correlation, $-1$ indicates a perfect negative correlation,
and $0$ indicates that the two sets of data are uncorrelated. In all
cases the results are $>0.65$, with most $>0.8$, indicating a
significant correlation between these SF diagnostics. Additionally,
the \halum\,--\radiolum\ relation appears to be better correlated than
\uvlum\,--\radiolum.

To fit the correlations, we perform a least squares fit weighted by
the errors in both variables to be correlated. We list the resulting
$\chi^2$, probability of $\chi^2$, and the fitted slopes and errors in
Table~\ref{fits_table}. We note there is a large scatter about these
best-fit lines (Figure~\ref{correlations}) -- with up to an order of
magnitude difference between the SFRs derived from different
diagnostics -- and systematic offsets from the lines of constant SFH.

As expected if the 1.4\,GHz fluxes reliably trace the SF free from
dust extinction, galaxies are under-luminous in H$\alpha$ and UV when
compared to the 1.4\,GHz luminosity. Also, there appears to be
luminosity dependencies in Figure~\ref{correlations} (most pronounced
in the UV/1.4\,GHz plot) in the sense that the brighter radio sources
are more under-luminous in UV, as shown by the slopes of the best-fit
lines. This effect is not seen to such a large extent in the
\halum\,--\,\radiolum\ plot, though the effect may still be present to
some degree. It is unclear whether this luminosity-dependent effect
reflects a non-linearity in the relationship between radio flux and SF
or a greater degree of extinction in more energetic systems.  This is
quantitatively similar to the situation found by
\citet{1998ApJ...507..155C} for a larger, although less homogeneous,
sample.

\begin{figure*}
\plotone{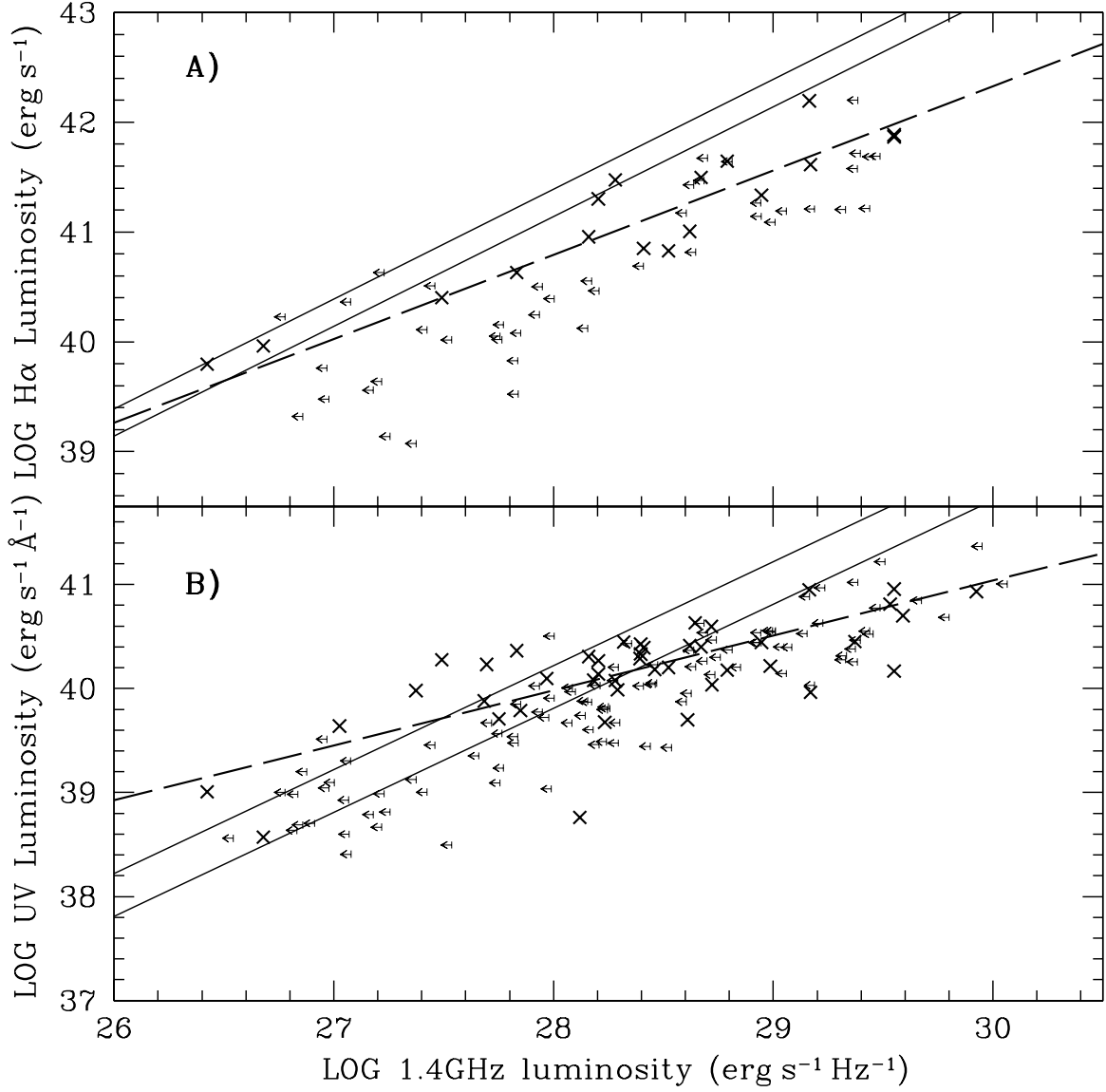}

\figcaption{As Fig.~\ref{correlations}, but with the H$\alpha$ and UV
  luminosities corrected for dust as in S2000 (based on the Balmer
  decrement), and with error bars removed for clarity.
\label{correlations_dust}}
\end{figure*}

The effect of dust on the H$\alpha$ and UV luminosities is a major
source of uncertainty. In our previous work (S2000), in galaxies where
both H$\alpha$ and H$\beta$ were detected, nebular emission lines were
corrected using the Balmer decrement, and these Balmer-derived
corrections then extended into the UV using a
\citet{2000ApJ...533..682C} reddening law \citep[see
also][]{1997uulh.conf..403C}. An average of these Balmer corrections
was then used in galaxies where H$\beta$ was not detected at an
adequate S/N. After corrections based on the Balmer decrement have
been applied (Figures~\ref{correlations_dust}a
and~\ref{correlations_dust}b), the SFRs derived from the different
diagnostics agree rather better.  However, the H$\alpha$/UV
luminosities are still generally under-luminous when compared to
1.4\,GHz luminosities, and the systematic effects and scatter seen
before dust correction remain (Table~\ref{fits_table}).

We investigate the significance of the scatter seen in these plots in
two ways. Firstly, we compare the residuals of the 1.4\,GHz
luminosities from the weighted best-fit lines with the measurement
errors in the radio luminosities. Whilst the residuals show a flat
distribution over the range 0\,--\,1.5 (in log luminosity), the
distribution of the errors is markedly different, being strongly
peaked in the range 0\,--\,0.4 (again in log-luminosity units). A
similar pattern is found for the dust-corrected H$\alpha$ and UV
measurements. Secondly, we note that the straight line fitting results
(Table~\ref{fits_table}) generally give \textit{poor} fits to the data
for the \radiolum-\uvlum\ relation (denoted by the $\chi^2$ values),
even though the datasets are undoubtedly correlated.  This implies
that either the observational errors are underestimated (a scenario we
do not believe to be true) or that the scatter is larger than would be
expected for a tight linear correlation. While the size of this sample
is small, making these kinds of tests difficult, we nonetheless
conclude that the scatter we see in these plots cannot arise purely
from observational errors in our data (see also
section~\ref{sec:effects-select-bias}).

We conclude that although the SF diagnostics correlate over several
orders of magnitude, there is a large (approximately an order of
magnitude) and statistically significant scatter around these
correlations which is not removed after simple dust corrections. This
implies that our best-fit correlations are not consistent with
calibrations based on constant SFHs, a tight, linear \radiolum\,--\,SFR
relationship, \textit{and} extinction corrections which are
independent of galaxy luminosity.

\section{Discussion}
\label{sec:discussion}

In this section, we will discuss the implications of the results of
Section~\ref{sec:observ-results} and try to resolve the apparent
discrepancies highlighted there. The major features of the dataset are
the offsets (and non-linearities) when compared to our
standard scenario, and the scatter that we see around the best-fit
lines of the correlations.

\subsection{The Contribution of Active Galactic Nuclei}
\label{sec:cont-non-starf}

We start with the hypothesis that the non-linearity observed in
Figure~\ref{correlations} is due to the inclusion of active galaxies,
or at least objects where SF is not the dominant emission mechanism at
radio frequencies (for example AGN, where the radio emission may be
dominated by a nuclear ``monster''). To investigate this, we search
for evidence of different populations in the spectra of the galaxies:
those with detected {\em narrow-line} H$\alpha$ emission (i.e.
\textit{known} to be star-forming) and those with weak or non-existent
H$\alpha$ emission.  We exclude those galaxies observed with the WIYN,
where the spectral wavelength coverage is insufficient to detect
H$\alpha$. Of the remainder, only one galaxy has no detectable
H$\alpha$ emission (formally a fraction of 4\%, compared with
$\sim30\%$ in the FOCA sample as a whole), and it lies away from the
remainder of the galaxies (marked on Figure~\ref{correlations}). We
see evidence that the 1.4\,GHz luminosity varies as a function of
$(UV-B)_0$ colour (with higher luminosity systems being bluer), and
note that the object with no H$\alpha$ is the reddest object in our
sample ($\simeq6.5$ c.f. a median $(UV-B)_0\simeq0$ for this sample).
This is consistent with the scenario that objects with little or no
H$\alpha$ emission are weakly star-forming or early-type galaxies,
possibly hosting AGNs, which are responsible for the observed UV and
1.4\,GHz luminosities.

No evidence is found for other significant AGN contamination (e.g.
broad emission lines or unusual emission line ratios) in those objects
with H$\alpha$ emission (see S2000 for a fuller discussion; see also
Contini et al. in preparation). We conclude that whilst there are
undoubtedly AGN and non star-forming galaxies present in our sample,
we find no evidence that they are responsible for the scatter and
non-linearities in the observed relations.

\subsection{Non-linearities in the diagnostics plots}
\label{sec:non-line-diagn}

One of most interesting results from this study is the slopes of the
best-fit relations, which are significantly different from those
expected, assuming constant SF scenarios (i.e. a slope of one, see
Section~\ref{sec:effects-select-bias}). One possibility is using the
\citet{2000ApJ...533..682C} extinction result to apply optical
extinction measures (the Balmer ratio) at UV wavelengths is not
appropriate for our sample of galaxies, for example if complex dust
geometries ensured a non-trivial relation between H$\alpha$ and UV
attenuations. Until some independent measure of the UV extinction is
available -- for example measures at other UV wavelengths in addition
to 2000\,{\AA} -- such a possibility must be deferred to a later
analysis. There appear to be two other possible explanations for these
observed non-linearities (as well as non-uniformities in the SFH which
we discuss in the next section); the related effect of a
luminosity-dependent dust correction, and a non-linearity in the
\radiolum\,--\,SFR calibration. We discuss each in turn.

Our previous dust corrections take little account of any possible
luminosity dependence, as only few galaxies posses H$\alpha$, H$\beta$
and 1.4\,GHz emission. Such luminosity effects could arise if
intensely star-forming galaxies (i.e. those with larger 1.4\,GHz
luminosities), or the star-forming regions of these galaxies, posses
dustier environments. \citet{1996ApJ...457..645W} investigated such
effects in a local sample of galaxies, and demonstrated that the
UV(2000\,{\AA})/FIR ratio decreases with increasing FIR luminosity,
implying that the dust opacity may increase in more strongly
star-forming environments.  If such effects were present in our
sample, then dust corrections which accounted for this would have the
effect of `rotating' our best-fit lines in an anti-clockwise
direction.

To investigate this in more detail, we follow
\citet{2001astro.ph..0103253} and attempt to derive a relation between
the intrinsic SFR in a galaxy and the extinction present. Ideally,
this could be done by examining the H$\alpha$/H$\beta$ trend with a SF
diagnostic unaffected by dust, e.g.  the 1.4\,GHz luminosity.
Unfortunately, the sample size of objects with radio, H$\alpha$ and H$\beta$
detections is currently so small that such a comparison is not
conclusive.

\begin{figure*}
  \plotone{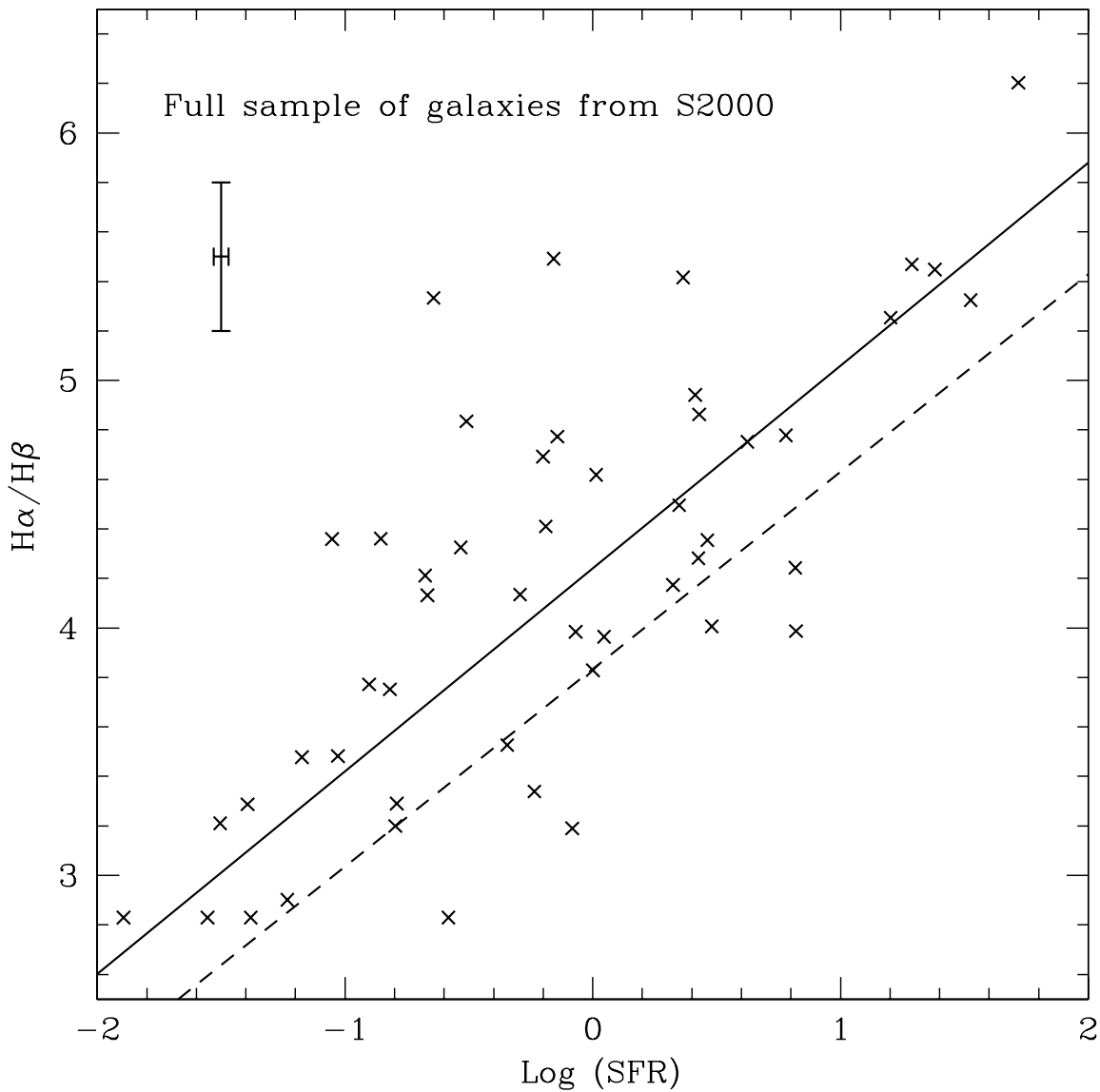} \figcaption{The ratio of H$\alpha$ to H$\beta$ as a
    function of SFR derived from dust corrected H$\alpha$
    luminosities. The points are from the full spectroscopic sample of
    S2000, regardless of whether they have a radio detection in this
    present survey. The solid line indicates the weighted
    least-squares best fit to the dataset; the dashed line shows the
    relationship derived by \citet{2001astro.ph..0103253} for an
    independent sample of galaxies. The error-bar in the top left
    corner demonstrates the median errors in the two parameters
    correlated.
\label{ratio_sfr}}
\end{figure*}

Instead, to explore the consequence of a luminosity dependence, we
appeal to the full sample of S2000 galaxies (whether or not they have
a radio detection in the present survey) and correlate the
H$\alpha$/H$\beta$ ratio with both \textit{uncorrected} and
\textit{Balmer-corrected} H$\alpha$ luminosity. We see a clear trend
in both cases, with intrinsically brighter H$\alpha$ luminosity
galaxies having a larger H$\alpha$/H$\beta$ ratio, even before dust
correction of the H$\alpha$. We demonstrate this relationship in
Figure~\ref{ratio_sfr}. It is important to realize that this
relationship has a large scatter (partly due to the uncertainties in
determining the H$\alpha$ and H$\beta$ fluxes -- see S2000), and so
such corrections should only be used in a statistical manner.

By fitting the corrected relationship (weighted by the errors in both
the Balmer ratio and the H$\alpha$ luminosity), we are able to
construct an observed relationship between the intrinsic SFR in an
object (as governed by the corrected H$\alpha$ emission) and the
Balmer decrement, which can then be used for a subsequent dust
correction.  This fitted relationship is

\begin{equation}
\frac{\rm{H}\alpha}{\rm{H}\beta}=0.82\times \log(SFR)+4.24.
\end{equation}

\noindent
and has a correlation coefficient of 0.65.
\citet{2001astro.ph..0103253} use this technique, but derive the SFRs
independently from FIR observations. This approach is obviously to be
preferred, as the FIR provides a measure of the intrinsic SFR
independently from the H$\alpha$ measure used above. They find
$\rm{H}\alpha/\rm{H}\beta=0.80\times\log(SFR)+3.83$, in remarkably
good agreement with the above estimate despite differences in sample
selection and the SF diagnostic used to determine the dust-corrected
SFR.  Therefore, for each object detected in our radio survey -- i.e.
with an independent measure of (presumed) dust-free SF at 1.4\,GHz --
we can correct the observed H$\alpha$ and UV luminosities using
equation~(6) and a \citet{2000ApJ...533..682C} law to extend to UV
wavelengths.  Obviously, such an approach can only correct the general
trend seen in our sample, and will not remove the galaxy to galaxy
scatter.

\begin{figure*}
 \plotone{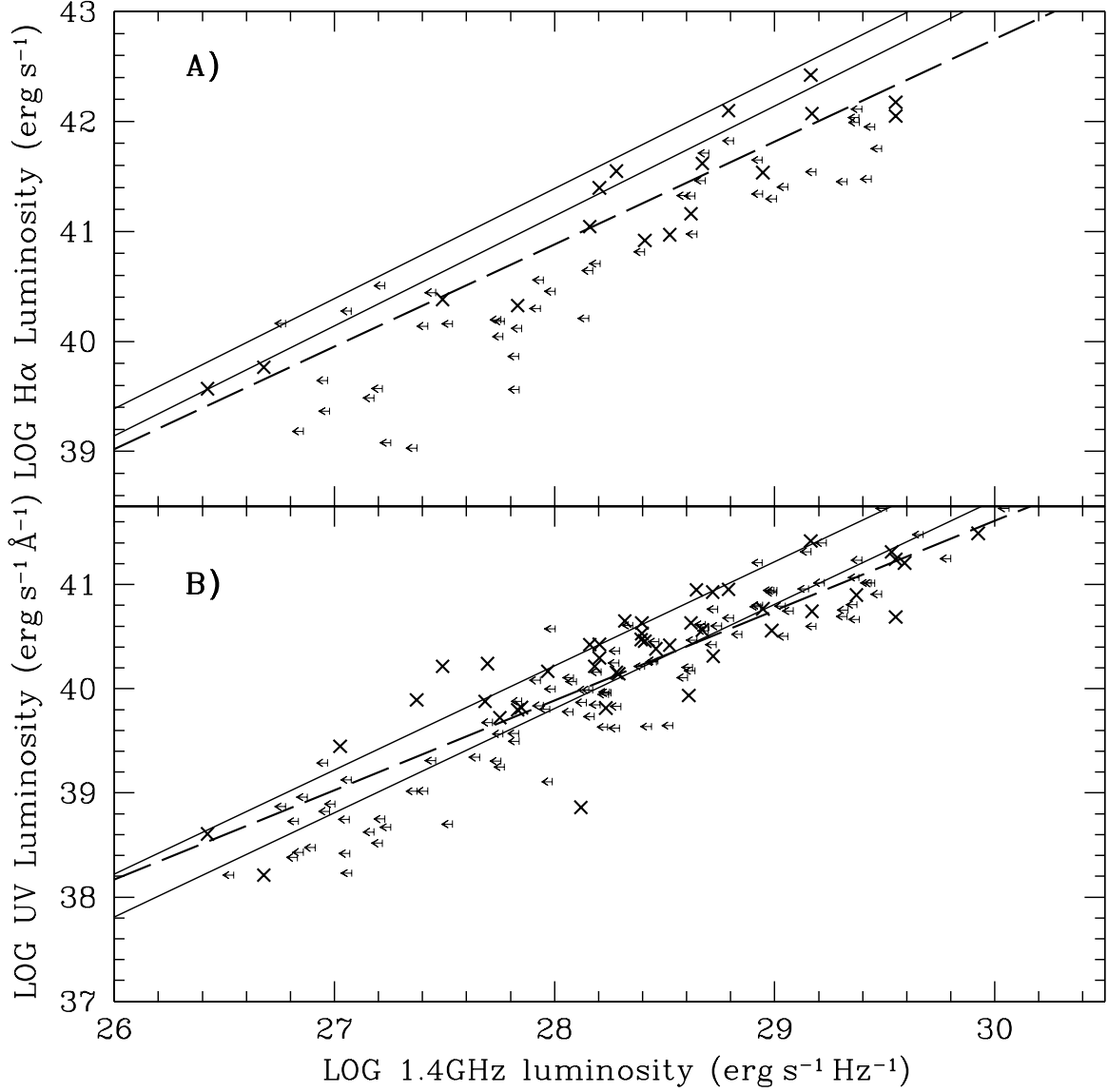}
  \figcaption{As Fig.~\ref{correlations}, but with the H$\alpha$ and
    UV luminosities corrected according to a luminosity dependent
    extinction prescription, again with error bars removed for clarity.
\label{correlations_lumdependent}}
\end{figure*}

The resulting correlations are shown in
Figure~\ref{correlations_lumdependent}. The slope of the best-fit line
is now closer to unity (predicted in constant SF scenarios and with
linear 1.4\,GHz to SFR conversions), though the UV/1.4\,GHz relation
is still slightly too shallow.  Adopting the
\citet{2001astro.ph..0103253} correction gives an almost identical
result. Larger and deeper samples are clearly needed with a more
comprehensive wavelength coverage, but this first analysis suggests
that empirical, luminosity-dependent extinction corrections can go
some way to explaining the slopes of our best-fit relations.

The second explanation considers the validity of the calibration of
the 1.4\,GHz luminosities and the conversion into a SFR. This
calibration is based on the observed conversion values of the
non-thermal radio luminosity to the supernova rate of our own galaxy.
If, in small (or low-SFR) galaxies, cosmic ray electrons were able to
escape more easily than in larger systems, or at least have a
different escape fraction compared to our galaxy, one may
underestimate the radio-derived SFR in low luminosity systems, and
possibly overestimate in high-luminosity systems
\citep[see for example][]{1990MNRAS.245..101C,1991ApJ...376...95C}. Such an effect
could generate a similar `luminosity-dependent' effect to that
observed.  A full analysis of this complex issue must await a more
complete dataset.

\subsection{The physical origin of the scatter}
\label{sec:phys-orig-scatt}

\begin{figure*}
  \plotone{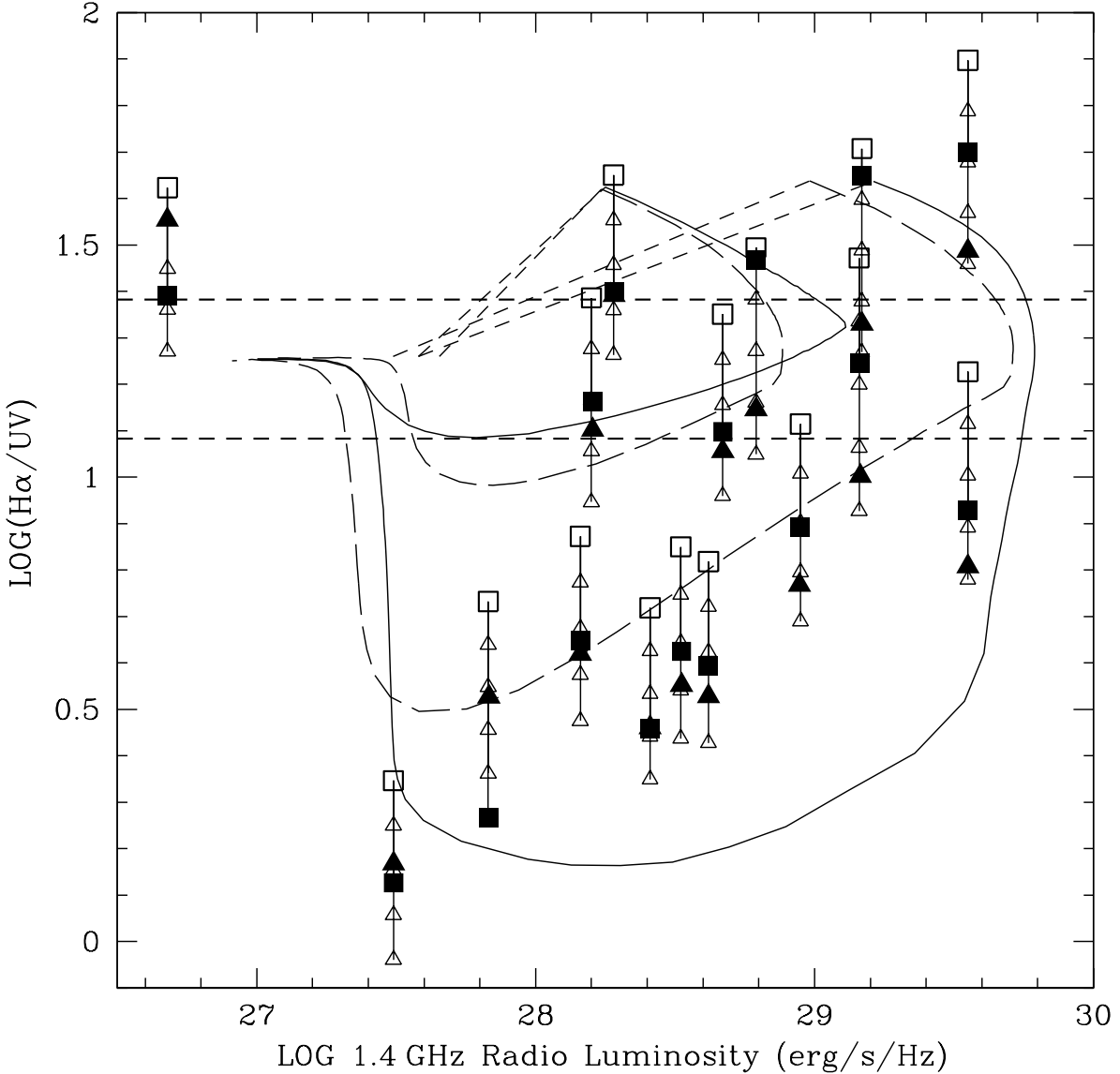} \figcaption{ The
    variation in the H$\alpha$/UV ratio with 1.4\,GHz luminosity for
    our sample galaxies. OPEN SQUARES: galaxies with no extinction
    corrections applied, FILLED SQUARES: Extinction corrections from
    S2000, FILLED TRIANGLES: Corrections based on a
    luminosity-dependent prescription, OPEN TRIANGLES: Extinctions
    with $A_V=0.5,1.0,1.5\;\&\; 2.0$. The horizontal dashed lines show
    the range of values obtained by varying the IMF upper mass cutoffs
    from 125\,$\rm{M_{\odot}}$ (top) to 50\,$\rm{M_{\odot}}$ (bottom).
    We also show predictions based on bursts of SF superimposed on
    exponential SFHs.  Here, the solid lines represent bursts of
    duration 110\,Myr (small loop) and 10\,Myr (large loop), mass 20\%
    in both cases.  The long dashed lines represent bursts of mass 5\%
    (small loop) and 35\% (large loop), duration 30\,Myr in both
    cases.  The short dashed lines are areas at the start of the burst
    where galaxies would spend an extremely small amount of time, they
    then move clockwise around the loops over
    time.\label{ratio_bursts}}
\end{figure*}

The second major feature in our data is the statistically significant
scatter around the best-fit lines. As we saw in
Section~\ref{sec:effects-select-bias}, generating this scatter via an
increased range of dust extinctions in our sample raises difficulties
in reconciling the radio-detected fraction of UV galaxies with those
expected in simulations (assuming a \citet{2000ApJ...533..682C} law).
We investigate this effect in several ways by examining the variation
of the H$\alpha$/UV ratio as a function of 1.4\,GHz luminosity
(Figure~\ref{ratio_bursts}). This diagnostic diagram is the least
sensitive to dust, affected only by {\em differential extinction}
between H$\alpha$ and the UV (assuming the radio data are dust free).
Also included on Figure 6 is the position of each galaxy, using
different extinction corrections, based on the Balmer decrement, the
luminosity-dependent corrections and the range of corrections
corresponding to visual extinctions of $A_V=0.5,1.0,1.5\;\&\;2.0$.
Again, we adopt the \citet{2000ApJ...533..682C} law to extend the
extinction corrections into the UV.

Figure~\ref{ratio_bursts} implies that, for a constant SFR, it is
unlikely that dust or IMF variations (i.e. varying the upper mass
cut-off) are sufficient to explain the scatter. Even considering the
full range of dust corrections, some galaxies have H$\alpha$/UV ratios
that cannot be reproduced.  We conclude that while a significant
fraction of the scatter in this diagram is due to different levels of
dust extinction among the galaxies, these dust corrections cannot (in
their present form) explain the entire dataset.  Other more complex
dust geometries, for example that of \citet{2000ApJ...539..718C} which
models the time-varying attenuation of H$\alpha$ line emission and UV
continuum, will again be deferred to a later paper.

In S2000, a comparison of H$\alpha$ and UV luminosities for the entire
FOCA sample revealed discrepancies that were difficult to reconcile
within the framework of simple dust corrections. Therefore, S2000
considered the possibility that a fraction of the FOCA galaxies were
undergoing bursts of short-term, intensive SF, which were able to
explain the observed dataset to a certain degree. Such scenarios
naturally generate a scatter on the H$\alpha$/UV/1.4\,GHz planes due
to differing dependencies on the timescale of SF of these different
diagnostics (see Section~\ref{sec:diagnostic-relations}).  For
example, UV continuum radiation is present after the completion of a
starburst for a longer duration than nebular or radio emission.  By
applying this hypothesis to the current dataset, we attempt to explain
the likely cause of the observed trends in Figure~\ref{correlations}
and \ref{correlations_dust}. Whilst this dataset is too small for a
thorough quantitative analysis, we can nevertheless test the
reliability of such approach. Therefore, we relax the assumption of a
constant SFH and consider a temporally varying SFH for our field
galaxies.  This is done by superimposing exponential starbursts of
varying strength and duration onto underlying exponential SFHs, as a
function of time, thereby simulating a `star-bursting + evolved
population' galaxy, from which we can use \pegase\ to estimate the
evolution of the H$\alpha$, UV, and 1.4\,GHz luminosities over time
(see Section~\ref{sec:diagnostic-relations}).

The effect of varying the burst parameters is demonstrated in
Figure~\ref{ratio_bursts} by assuming different bursts corresponding
to over reasonable ranges (5--35\% and 10--110\,Myr duration).  The
bursts occur at a galactic age of 6\,Gyr. For illustrative purposes,
each artificial galaxy has a mass of $10^{10}\,\rm{M}_{\odot}$, of course
in reality this is unlikely to be true, but we must await near-IR data
to explore this further. It is clear that all of the scatter could be
explained in terms of a temporally varying SFHs using bursts of
different parameters, though with this small sample size it is not
possible to constrain the burst parameters in any meaningful manner.

\section{Conclusions}
\label{sec:conclusions}

We have presented the first results from a decimetric radio survey of
nearby galaxies, with the ultimate goal of comparing SF diagnostics
for a homogeneous sample of star-forming galaxies. We find broad
correlations over several orders of magnitude between the different SF
diagnostics but with a large galaxy-to-galaxy scatter and
offsets/non-linearities from relations predicted using simple dust
extinction and SF scenarios. By dividing  our sample into two (those
with and without {\em detected} H$\alpha$ emission), we tentatively
conclude that the scatter and offsets that we see are not due to a
significant non-starforming population of galaxies.

We find evidence for luminosity-dependent effects in our dataset, and
show that luminosity dependent dust corrections or a mis-calibration
of the 1.4\,GHz-SFR calibration, or a combination of both, can go some
way to resolving this effect. We also demonstrate that, over realistic
ranges, differential extinction between galaxies cannot be solely
responsible for the scatter in our relations; indeed our dataset
argues against significant extinction in our sample. We conclude the
discrepancies between different SF diagnostics can only be partly
explained by simple models of dust extinction in galaxies. These
models cannot by themselves explain all the observed differences,
introducing the need for temporally varying SFHs and/or more complex
models of extinction, to explain the entire dataset.

\acknowledgments{We thank the anonymous referee for detailed comments
  which improved this manuscript. The National Radio Astronomy
  Observatory is a facility of the National Science Foundation
  operated under cooperative agreement by Associated Universities,
  Inc. The William Herschel Telescope is operated on the island of La
  Palma by the Isaac Newton Group in the Spanish Observatorio del
  Roque de los Muchachos of the Instituto de Astrofisica de Canarias.}


\begin{deluxetable}{ccccccc}
\tabletypesize{\scriptsize} 
\tablecaption{The details of the different samples in our combined survey.\label{sample_numbers}}
\tablehead{\colhead{Sample} & \colhead{Total} & \colhead{Total} & \colhead{Total} & \colhead{Total} & \colhead{Total} & \colhead{Total}\\ & \colhead{(\vsad)} & \colhead{(\imfit)} & \colhead{(pixel-sum)} & & \colhead{(with H$\alpha$)\tablenotemark{2}} & \colhead{(with fluxed H$\alpha$)\tablenotemark{3}}}
\startdata
Largest sample                         & 26 & 25 & 3& \textbf{54} & 30 & (21)\\
(minus $>1$ POSS counterparts)         & 23 & 21 & 3 & \textbf{47}& 27 & (19)\\
(minus Coma galaxies)\tablenotemark{1} & 22 & 18 & 3 & \textbf{43}& 25 & (17)\\
\tableline
\enddata
\tablenotetext{1}{Corresponding to the default sample used in the analysis}
\tablenotetext{2}{Excludes WIYN galaxies with insufficient spectral coverage to detect H$\alpha$}
\tablenotetext{3}{WHT observed galaxies with adequate flux calibration}
\end{deluxetable}

\begin{deluxetable}{ccccccc}
\tablecaption{Factors derived to convert diagnostic luminosities into
  SFRs for various luminosities and
  timescales, in the sense SFR(M$_{\odot}$\,yr$^{-1}$)=Luminosity/conversion factor.\label{conversion_factors}}
\tablehead{\colhead{Z} & \colhead{H$\alpha$} & \multicolumn{3}{c}{UV ($\rm{erg\,s^{-1}\,{\AA}^{-1}}$)} & \multicolumn{2}{c}{1.4\,GHz ($\rm{erg\,s^{-1}}$)} \\
 \colhead{} & \colhead{($\rm{erg\,s}^{-1}$)} & \colhead{10\,Myr} & \colhead{100\,Myr} & \colhead{1000\,Myr}& \colhead{10\,Myr} & \colhead{100\,Myr}}
\startdata
0.0004 & 2.02e$^{41}$ & 3.24e$^{39}$ & 5.87e$^{39}$ & 7.68e$^{39}$ & 2.03e$^{27}$ & 8.84e$^{27}$\\
0.004 & 1.67e$^{41}$ & 3.39e$^{39}$ & 5.84e$^{39}$ & 7.24e$^{39}$ & 2.06e$^{27}$ & 1.08e$^{28}$\\
\bf 0.02\tablenotemark{1} & \bf 1.22e$^{41}$ & \bf 3.75e$^{39}$ & \bf 5.76e$^{39}$ & \bf 6.56e$^{39}$ & \bf 2.39e$^{27}$ & \bf 8.85e$^{27}$\\
0.05 & 0.87e$^{41}$ & 3.86e$^{39}$ & 5.49e$^{39}$ & 5.95e$^{39}$ & 3.28e$^{27}$ & 8.78e$^{27}$\\
\tableline
\tablenotetext{1}{Corresponding to solar metallicity and the default conversion factors used in the analysis}
\enddata
\end{deluxetable}

\begin{deluxetable}{cccccccc}
\tabletypesize{\scriptsize}
\tablecaption{The Diagnostic Correlations\label{fits_table}}
\tablehead{\colhead{Relation} & \colhead{Dust} & \colhead{Number of} & \colhead{Correlation} & \colhead{Spearman} & \colhead{$\chi^2$} & \colhead{$\chi^2$} & \colhead{Slope}\\
\colhead{} & \colhead{Correction} & \colhead{Sources} & \colhead{Coeff.\tablenotemark{1}} & \colhead{Coeff.\tablenotemark{1}} & \colhead{} & \colhead{Probability\tablenotemark{2}} & \colhead{}}
\startdata
H$\alpha$-radio & None             & 17 & 0.908 & 0.873 & 38.9  & 0.0007    & 0.73 $\pm$ 0.07\\
UV-radio        & None             & 43 & 0.752 & 0.745 & 101.4 & $<0.0001$ & 0.58 $\pm$ 0.05\\
H$\alpha$-radio & Balmer decrement & 17 & 0.927 & 0.900 & 20.5  & 0.155     & 0.77 $\pm$ 0.06\\
UV-radio        & Balmer decrement & 43 & 0.711 & 0.634 & 136.2 & $<0.0001$ & 0.53 $\pm$ 0.05\\
H$\alpha$-radio & Lum-dependent    & 17 & 0.936 & 0.884 & 33.4  & 0.0024    & 0.93 $\pm$ 0.07\\
UV-radio        & Lum-dependent    & 43 & 0.865 & 0.877 & 84.1  & $<0.0001$ & 0.85 $\pm$ 0.05\\
\enddata
\tablenotetext{1}{Where 1 equals a perfect positive correlation, 0 indicates no correlation}
\tablenotetext{2}{Where smaller values indicate poorer fits}
\end{deluxetable}

\end{document}